\documentclass[final]{elsart}

\journal{Science of Computer Programming}
\date{}

\mathchardef\Omega="700A

\usepackage{amsmath}
\usepackage{amssymb}
\usepackage{stmaryrd}
\usepackage{url}
\usepackage{verbatim}
\usepackage{ifthen}
\usepackage{varioref}
\usepackage{hhline}
\usepackage{dcolumn}
\usepackage{longtable}
\usepackage{pstricks,pst-node,pst-plot}

\usepackage[italian,english]{babel}

\newcommand{\italian}[1]{\foreignlanguage{italian}{#1}}

\usepackage{listings}
\newcolumntype{.}{D{.}{.}{-1}}

\newcommand*{\Cplusplus}{{C\nolinebreak[4]\hspace{-.05em}\raisebox{.4ex}{\tiny\bf ++}}}

\newcommand{\st}{\mathrel{.}}

\providecommand*{\Zset}{\mathbb{Z}}            

\providecommand*{\Rset}{\mathbb{R}}            

\begin{document}


\begin{frontmatter}

\title{The Parma Polyhedra Library: \\ Toward a Complete Set of Numerical
Abstractions for the Analysis and Verification
of Hardware and Software Systems\thanksref{th}}

\thanks[th]{This
work has been partly supported by PRIN project
``AIDA: Abstract Interpretation Design and Applications.''
}

\author[Parma]{Roberto Bagnara},
\ead{bagnara@cs.unipr.it}
\author[Leeds]{Patricia M. Hill},
\ead{hill@comp.leeds.ac.uk}
\author[Parma]{Enea Zaffanella}
\ead{zaffanella@cs.unipr.it}

\address[Parma]{Department of Mathematics, University of Parma, Italy}
\address[Leeds]{School of Computing, University of Leeds, UK}

\begin{abstract}
Since its inception as a student project in 2001,
initially just for the handling (as the name implies) of convex polyhedra,
the \emph{Parma Polyhedra Library} has been continuously improved and
extended by joining scrupulous research on the theoretical foundations
of (possibly non-convex) numerical abstractions to a total adherence
to the best available practices in software development.
Even though it is still not fully mature and functionally complete,
the Parma Polyhedra Library already offers a combination of
functionality, reliability, usability and performance that is not
matched by similar, freely available libraries.
In this paper, we present the main features of the current version of
the library, emphasizing those that distinguish it from other similar
libraries and those that are important for applications in the field
of analysis and verification of hardware and software systems.
\end{abstract}


\end{frontmatter}

\section{Introduction}

The \emph{Parma Polyhedra Library} (PPL) is a collaborative project
started in January~2001 at the Department of Mathematics
of the University of Parma, Italy.
Since 2002, the library is actively being developed also at
the School of Computing of the University of Leeds, United Kingdom.
The PPL, that was initially limited ---as the name implies---
to the handling of (not necessarily topologically closed)
convex polyhedra \cite{BagnaraRZH02}, is now aimed at becoming
a truly professional, functionally complete library for the handling
of numeric approximations targeted at abstract interpretation and
computer-aided verification of hardware and software systems.

In this paper we briefly describe the main features of the PPL.
Unless otherwise stated, the description refers to version 0.9 of the
library, released on March 12, 2006.
In other cases the described features have not yet been included into
an official release, but are still available from the PPL's public
CVS repository (since development of the library takes place on that
repository, the PPL is always in a state of continuous release).

In the sequel we will compare the PPL with other libraries
for the manipulation of convex polyhedra.
These are:\footnote{We restrict ourselves to those libraries that are
freely available and provide the services required by applications
in static analysis and computer-aided verification.}
\begin{itemize}
\item
\emph{PolyLib} (version 5.22.1),
originally designed by D.~K.~Wilde \cite{Wilde93th},
based on an efficient C implementation of N.~V.~Chernikova's
algorithm \cite{Chernikova64,Chernikova65,Chernikova68}
written by H.~Le~Verge \cite{LeVerge92}, and further developed
and maintained by V.~Loechner
\cite{Loechner99};\footnote{\url{http://icps.u-strasbg.fr/~loechner/polylib/}.}
\item
\emph{New Polka} (version 2.1.0a), by B.~Jeannet
\cite{NEW-POLKA-1-1-3c};\footnote{\url{http://pop-art.inrialpes.fr/people/bjeannet/newpolka/index.html}.}
\item
the polyhedra library that comes with the \emph{HyTech} tool
(version 1.04f) \cite{HenzingerHW97b};\footnote{\url{http://embedded.eecs.berkeley.edu/research/hytech/}.}
\item
the \emph{Octagon Abstract Domain Library} (version 0.9.8), by A.~Min\'e
\cite{Mine01b,Mine05th}.\footnote{\url{http://www.di.ens.fr/~mine/oct/}.}
\end{itemize}

For reasons of space and opportunity, the paper concentrates on
introducing the library to prospective users.  As a consequence, we
assume the reader has some familiarity with the applications of
numerical abstractions in formal analysis and verification methods and
that she is not immediately concerned by the theory that lies
behind those abstractions.  However, the paper provides a complete
set of references that enable any interested reader to reconstruct
every detail concerning the library, the applications and the
relevant theory.

The paper is structured as follows:
Section~\ref{sec:abstractions} introduces the numerical abstractions
currently supported by the Parma Polyhedra Library;
Section~\ref{sec:features} describes, also by means of examples, some
of the most important features of the library;
Section~\ref{sec:efficiency} gives some indications concerning efficiency;
Section~\ref{sec:development-plans} illustrates the current development plans
for the library;
Section~\ref{sec:discussion} reviews some of the applications using
the PPL and concludes.

\section{Currently Supported Abstractions}
\label{sec:abstractions}

The numerical abstractions currently supported by the Parma Polyhedra
Library are the domains of polyhedra, bounded difference shapes,
octagonal shapes and grids; powersets of these using the generic
powerset construction; and linear programming problems.
For each of the supported abstractions, the library features all the
operators required by semantic constructions based on
\emph{abstract interpretation} \cite{CousotC77,CousotC79};
these are summarized in the following paragraphs.

\paragraph*{Construction}
The domain elements can be initialized by means of a variety of
constructors; in particular, they can be defined to be a vector space
for a specified number of dimensions ---each dimension representing
some concrete entity relevant for the particular analysis---
indicating that at this stage nothing is known about the entities, or
the empty abstraction (again for a given number of dimensions),
describing an inconsistent (unreachable) state.
An element can also be initialized by means of \emph{constraints},
specifying conditions its points must satisfy; alternatively, it can
be described by \emph{generators}, that are meant to be parametrically
combined so as to describe all of its points.
New elements can also be constructed by copying an existing element in
the same abstraction.

\paragraph*{Refinement and Expansion}
Any domain element can be refined by the addition of further constraints
its points must satisfy, thereby improving the precision of the description:
a typical application of refinement is to model the effect of
conditional guards in if-then-else and while-loop statements.
The expansion operators allow users to add new generators
so as to expand the set of points the element must contain;
these may be used, for instance, to ``forget'' some information
regarding a space dimension, so as to model arbitrary input by the user.

\paragraph*{Upper and Lower Bounds}
It is often useful to compute an upper or lower bound of two domain
elements so as to obtain a correct approximation of their union or
intersection.
For example, an analyzer could use an upper bound operator
to combine approximations
that have been computed along the alternative branches of an
if-then-else statement while
lower bound operators are needed, in combination with conversion operators
(see below), when conjunctively merging different approximations computed
along the same set of computation paths.

\paragraph*{Affine Images and Preimages}
A common form of statement in an imperative language
is the assignment of an affine expression to a program variable.
Their semantics can be modeled by
images (in a forward analysis) and/or
preimages (in a backward analysis) of affine transformations
on domain elements.
All domains fully support the efficient (and possibly approximate)
computation of affine images and preimages.
Also available are generalizations of these operators,
useful for approximating more complex assignment statements
(e.g., to model constrained nondeterministic assignments or
when a non-linear expression can be bounded from below and/or above
by affine expressions).

\paragraph*{Changing Dimensions}
An analyzer needs to be able to add, remove, and more generally reorganize
the space dimensions that are associated with the values of the concrete
entities it is approximating. The simple addition and removal of dimensions
is often needed at the entry and exit points of various kinds of
programming contexts, such as declaration blocks where concrete entities
(modeled by some of the space dimensions) may be local to the block.
More complex operators are useful to support the integration of the
results computed using different abstractions, that typically provide
information about different sets of concrete entities.
In the simplest case, when the abstractions are of the same kind but
provide information about disjoint sets of concrete entities, it is
enough to \emph{concatenate} the two abstract elements into a new one.
In more complex cases, when the described sets of concrete entities
have an overlap, the space dimensions of one of the abstract elements
need to be reconciled with those of the other, allowing for a correct
integration of the information.
This can be obtained by efficiently \emph{mapping} the space dimensions
according to a (partial) injective function.
The library also supports the \emph{folding} (and unfolding) of space
dimensions, which is needed to correctly and efficiently
summarize the information regarding collections of concrete
entities~\cite{GopanDMDRS04}.

\paragraph*{Conversion}
Non-trivial analyses are typically based on a combination of domains.
It is therefore important to be able to safely and efficiently convert
between different abstractions. These conversions enable, for
instance, the combination of domain elements
representing different kinds of information,
implementing so-called \emph{reduction} operators. Another important
application is the dynamic control of the precision/efficiency ratio:
in order to gain efficiency in the analysis of a specific context,
an element of a relatively precise domain can temporarily be converted into
an element of a weaker domain and then back to the stronger
abstraction on exit from that context.

\paragraph*{Comparison}
The analysis of a program fragment is implemented by computing an
over-approximation of its semantics.  Since the latter is usually
modeled as the least fixpoint of a continuous operator, a safe
analysis typically needs to iteratively compute an over-approximation
of this fixpoint.  Therefore, a suitable lattice-theoretic comparison
operator is needed so as to check for the convergence of the analysis.
Comparison operators are also useful in selected contexts so as to
efficiently predict whether or not some precision improving techniques
(e.g., reductions, widening variants and so on) are applicable.  For
all of the reasons above, each domain provides three different
comparison operators checking, respectively, equality, containment and
strict containment.

\paragraph*{Widening}
Most of the domains supported by the library admit \emph{infinite
ascending chains}.  These are infinite sequences of domain elements
such that every element in the sequence is contained in the element
that follows it. With these characteristics,
the fixpoint computations upon which abstract interpretation analysis
techniques are based could be non-terminating.
For this reason, the domains can be so employed
only in conjunction with appropriate mechanisms for enforcing and/or
accelerating the convergence of fixpoint computations: widening
operators \cite{CousotC76,CousotC77,CousotC92fr,CousotC92plilp}
provide a simple and general characterization for such
mechanisms.%
\footnote{Operators that do not provide a strict convergence guarantee
are more properly called \emph{extrapolation} operators.}
The PPL offers also several variations of the available widenings:%
\footnote{Some of these variations are extrapolation operators,
as the guarantee of convergence is conditional on the way they are used.}
widening ``with tokens'' (an improvement to the widening delay technique
proposed in \cite{Cousot81});  and widening ``up to''
\cite{Halbwachs93,HalbwachsPR97} (a technique whereby constraints that
are judged to be important by the application can be preserved from the
action of the widening).

\paragraph*{Other Operators}
The library offers many other operators for use in a variety of
more specialized contexts.
For instance, before performing a non-trivial domain combination,
an analyzer may need information
about a particular domain element (such as, checking whether it denotes
the empty set or the whole vector space; the
dimension of the vector space; the affine dimension of the element;
its relation with respect to a given constraint or generator, and so
on). Another operator supported by the library is the
\emph{difference} operator, which computes the smallest domain element
that contains the set difference of two elements; this is exploited in the
implementation of the finite powerset widening operator proposed
in~\cite{BagnaraHZ06STTT}. The library also provides
the \emph{time-elapse} operator used to model hybrid systems
\cite{HalbwachsPR97}.

The following sections briefly describe each of the supported domains.
The emphasis here is on the features that are unique to the PPL.  The
reader is referred to the cited literature and to the library's
documentation \cite{PPL-DEVREF-0-9,PPL-USER-0-9} for all the details.

\subsection{Closed and Not Necessarily Closed Polyhedra}

\begin{figure}
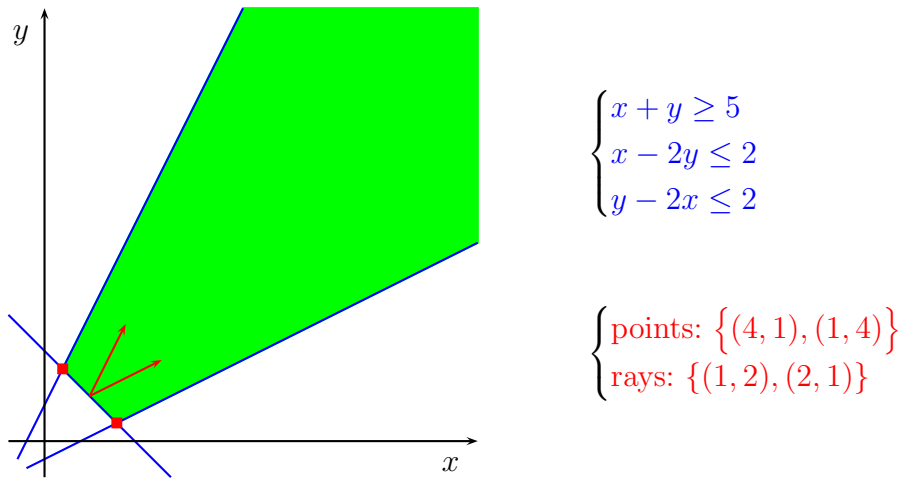

\centering
\begin{minipage}{8cm}
\[
\psset{xunit=2.4mm,yunit=2.4mm,runit=2.4mm}
\psset{origin={-0.5,-0.5}}
\pspicture*[](-2,-2)(25,25)
\pspolygon[fillstyle=solid,fillcolor=green,linecolor=green](1,4)(11,24)(24,24)(24,11)(4,1)
\psline[linecolor=blue](-2,7)(7,-2)
\psline[linecolor=blue](-1.5,-1)(11,24)
\psline[linecolor=blue](-1,-1.5)(24,11)
\psdots*[linecolor=red,dotstyle=square*,dotsize=0.6 0.6](1,4)(4,1)
\psline[linecolor=red]{->}(2.5,2.5)(4.5,6.5)
\psline[linecolor=red]{->}(2.5,2.5)(6.5,4.5)
\psline{->}(-2,0)(24,0)
\psline{->}(0,-2)(0,24)
\rput(23,-0.8){$x$}
\rput(-0.8,23){$y$}
\endpspicture
\]
\end{minipage}%
\begin{minipage}{5.7cm}
\begin{align*}
&\begin{cases}
  \blue{x + y \geq 5} \\
  \blue{x - 2y \leq 2} \\
  \blue{y - 2x \leq 2}
\end{cases} \\[1cm]
&\begin{cases}
  \red{\text{points: } \bigl\{ (4, 1), (1, 4) \bigr\}} \\
  \red{\text{rays: } \{ (1, 2), (2, 1) \}}
\end{cases}
\end{align*}
\end{minipage}
\caption{The double description method for polyhedra}
\label{fig:polyhedron-double-description}
\end{figure}

The Parma Polyhedra Library supports computations on the abstract
domain of convex polyhedra \cite{CousotH78,Halbwachs79th}.
The PPL implements both the abstract domain of
topologically \emph{closed convex polyhedra} (briefly called \emph{C polyhedra}
and implemented by class \verb+C_Polyhedron+)
and the abstract domain of \emph{not necessarily closed convex polyhedra}
(\emph{NNC polyhedra} for short, class \verb+NNC_Polyhedron+).
In both cases, polyhedra are represented and manipulated using the
\emph{Double Description} (DD) method of Motzkin et al.\ \cite{MotzkinRTT53}.
In this approach, a closed convex polyhedron can be specified in two ways,
using a \emph{constraint system} (class \verb+Constraint_System+)
or a \emph{generator system} (class \verb+Generator_System+):
the constraint system is a finite set of linear equality or inequality
constraints (class \verb+Constraint+);
the generator system is a finite set of different kinds of vectors,
collectively called \emph{generators},
which are rays and points of the polyhedron (class \verb+Generator+).
An example of double description is depicted
in Figure~\ref{fig:polyhedron-double-description}: the polyhedron
represented by the shaded region can be represented by the set of vectors
satisfying the constraints or, equivalently, by the set
\[
  \{\,
    \pi_1 \mathbf{p}_1 + \pi_2 \mathbf{p}_2
      + \rho_1 \mathbf{r}_1 + \rho_2 \mathbf{r}_2 \in \Rset^2
  \mid
    \pi_1, \pi_2, \rho_1, \rho_2 \in \Rset_+, \pi_1 + \pi_2 = 1
  \,\},
\]
where
$\mathbf{p}_1 = \left( \begin{smallmatrix}4 \\1\end{smallmatrix} \right)$,
$\mathbf{p}_2 = \left( \begin{smallmatrix}1 \\4\end{smallmatrix} \right)$,
$\mathbf{r}_1 = \left( \begin{smallmatrix}1 \\2\end{smallmatrix} \right)$,
$\mathbf{r}_2 = \left( \begin{smallmatrix}2 \\1\end{smallmatrix} \right)$,
and $\Rset_+$ is the set of non-negative real numbers.
In words, each vector can be obtained by adding a non-negative combination
of the rays and a convex combination of the points.

Implementation of convex polyhedra using the DD method offer some important
advantages to analysis and verification applications.
The first one is due to the ``mix'' of operations such applications require:
some of them are more efficiently performed on the representation with
constraints.  This is the case for the addition of constraints and for
the intersection, which is simply implemented as the union of constraint
systems, and for deciding whether a generator is subsumed or not by a
polyhedron (e.g., to decide whether a point is inside or outside).
Some operations are instead more efficiently performed on generators:
computing the convex polyhedral hull (just by taking the union of generator
systems), adding individual generators
(e.g., the addition of the rays
$\mathbf{r} = \left( \begin{smallmatrix}1 \\0\end{smallmatrix} \right)$ and
$-\mathbf{r}$ to the set of rays for the polyhedron in
Figure~\ref{fig:polyhedron-double-description}
is the easiest way to ``forget'' all the information concerning the
space dimension~$x$),
projection onto designated dimensions, deciding whether
the space defined by a constraint is disjoint, intersects or includes
a given polyhedron, finiteness/boundedness tests
(a polyhedron is finite/bounded if and only if it has no rays among its
generators),
and the \emph{time-elapse} operator
of~\cite{HalbwachsPR94,HalbwachsPR97}.
There are also important operations, such as the inclusion and equality tests
and the widenings, that are more efficiently performed
when \emph{both} representations are available.
Systems of constraints and generators enjoy a quite strong
and useful duality property.
Very roughly speaking, the constraints of a polyhedron are (almost)
the generators of the \emph{polar} \cite{NemhauserW88,StoerW70} of the
polyhedron, the generators of a polyhedron are (almost) the
constraints of the polar of the polyhedron, and the polar of the polar
of a polyhedron is the polyhedron itself.
This implies that computing constraints from generators is the same
problem as computing generators from constraints.
The algorithm of N.~V.~Chernikova \cite{Chernikova64,Chernikova65,Chernikova68}
(later improved by H.~Le~Verge \cite{LeVerge92}
and by K.~Fukuda and A.~Prodon \cite{FukudaP96})
solves both problems yielding a minimized system and can be implemented
so that the source system is also minimized in the process.
This is basically the algorithm employed by PolyLib, New Polka
and the Parma Polyhedra Library.
It is worth noticing that it is not restrictive to express the coefficients
of constraints and rays by integer numbers, as using rational numbers
would not result in increased expressivity (but would have a negative
impact on efficiency).  For points, a common integer denominator suffices.%
\footnote{The correctness requirements of applications in
the field of system's analysis and verification prevent the adoption
of floating-point coefficients, since any rounding error on the wrong
side can invalidate the overall computation.  For domains as
complicated as that of polyhedra, the correct, precise and reasonably
efficient handling of floating-point rounding errors is an open
issue.}

Restricting the attention to convex polyhedra, two of the main innovations
introduced by the PPL are the complete handling of NNC polyhedra and the
introduction of a new widening operator.
Apart from the PPL, the only libraries---among those that provide the
services required by applications in static analysis and computer-aided
verification---that support NNC polyhedra are the already mentioned New Polka
and the library by N.~Halbwachs, A.~Kerbrat and Y.-E.~Proy called, simply,
\emph{Polka} \cite{HalbwachsKP95}.  The Polka library, however, is not
available in source format and binaries are distributed under rather
restrictive conditions (until about the year 1996 they could be freely
downloaded), so our knowledge of it is as given in \cite{HalbwachsKP95},
the programmer's manual in a package that includes the actual library.
The support provided by Polka and New Polka for NNC polyhedra is
incomplete, incurs avoidable inefficiencies and leaves the client
application with the non-trivial task of a correct interpretation of
the results.  In particular, even though an NNC polyhedron can be
described by using constraint systems that may contain strict inequalities,
the Polka and New Polka libraries lack a corresponding extension
for generator systems.
In contrast, the PPL implements the proposal put forward in
\cite{BagnaraHZ05FAC}, whereby the introduction of \emph{closure points}
as a new kind of generator, allows a clean user interface,
symmetric to the constraint system interface for NNC polyhedra,
that is decoupled from the implementation.
As explained in detail in \cite{BagnaraHZ05FAC}, a user of the PPL
is fully shielded from implementation details such as the extra
$\epsilon$ dimension that users of the other libraries have to carefully
take into account.  Another feature that is unique to the PPL is
the support for the minimization of the descriptions of an NNC polyhedron:
we refer the interested reader to \cite{BagnaraHZ05FAC} for a precise
account of the impact this new feature has on performance and usability.

The original widening operator proposed by Cousot and Halbwachs
\cite{CousotH78,Halbwachs79th} is termed \emph{standard widening} since,
for 25 years, all analysis and verification tools that employed convex
polyhedra also employed that operator.
Nonetheless, there was an unfulfilled demand for more precise widening
operators.  The Parma Polyhedra Library, besides the standard
widening, offers the new widening proposed in \cite{BagnaraHRZ05SCP}:
on a single application this is always more precise than the standard
widening.  As these widenings are not monotonic, increased precision
on a single application does not imply increased precision on the
final result of the analysis.  In practice, however, an overall
increase of precision is almost always achieved
\cite{BagnaraHRZ05SCP}.

Both widenings can be improved, as said before, by applying the
``widening with tokens'' delay strategy or the ``widening up-to''
technique; moreover, ``bounded'' extrapolation operators are available
that provide additional precision guarantees over the widenings upon
which they are built.

\subsection{Bounded Difference Shapes}
\label{sec:bounded-difference-shapes}

By restricting to particular subclasses of linear constraints, it is
possible to obtain domains that are simpler and computationally more
efficient than the one of convex polyhedra.
One possibility, which has a long tradition in computer
science~\cite{Bellman57}, is to only consider
\emph{potential constraints}, also known as \emph{bounded differences}:
these are restricted to take the form $v_i - v_j \leq d$ or $\pm v_i \leq d$,
where $v_i$ and $v_j$ are variables and $d$, the \emph{inhomogeneous term},
belongs to some computable number family.
Systems of bounded differences have been used
by the artificial intelligence community as a way to reason about
temporal quantities \cite{AllenK85,Davis87},
as well as by the model checking community as an
efficient yet precise way to model and propagate timing requirements
during the verification of various kinds of concurrent systems
\cite{Dill89,LarsenLPY97}.
In the abstract interpretation field, the idea of using an abstract
domain of bounded differences was put forward in \cite{Bagnara97th}
and the first fully developed application of bounded differences
in this field can be found in~\cite{ShahamKS00}.
Possible representations for finite systems of bounded differences are
matrix-like data structures called \emph{difference-bound matrices}
(DBM) \cite{Bellman57} and weighted graphs called \emph{constraint networks}
\cite{Davis87}.  These representations, however, have a ``syntactic'' nature:
they encode sets of constraints rather than geometric shapes.
As pointed out in \cite{BagnaraHMZ05} this nature has several drawbacks,
the most important one being that natural extrapolation operators
do not provide a convergence guarantee, that is, they are not widenings.
This results into an extra burden on the client application, which has
to take into account the implementation details and use the domain elements
with care so as to avoid non-termination of the analysis.

In order to overcome the difficulties mentioned above and to continue
pursuing a complete separation between interface (which must be natural
and easy to use) and implementation (which must be efficient and robust),
the Parma Polyhedra Library offers the ``semantic'' domain of
\emph{bounded difference shapes}.
A bounded difference shape is nothing but a geometric shape, that is,
a convex polyhedron: its internal representation needs not concern
(and, in fact, is completely hidden from) the client application.
The class template implementing this domain in the PPL is
\verb+BD_Shape<T>+, where the class template type parameter \verb+T+
defines the family of numbers that are used to (correctly) approximate
the inhomogeneous terms of bounded differences.
The value of \verb+T+ may be one of the following:
\begin{itemize}
\item
a bounded precision native integer type (that is,
from \verb+signed+ \verb+char+ to \verb+long+ \verb+long+
and from \verb+int8_t+ to \verb+int64_t+);
\item
a bounded precision floating point type (\verb+float+, \verb+double+
or \verb+long+ \verb+double+);
\item
an unbounded integer or rational type, as provided by GMP
(\verb+mpz_class+ or \verb+mpq_class+).
\end{itemize}
Among other things, PPL's \verb+BD_Shape<T>+ offers the proper widening
operator defined in \cite{BagnaraHMZ05} and a user interface that matches
the interfaces of the general polyhedra classes \verb+C_Polyhedron+
and \verb+NNC_Polyhedron+.

\subsection{Octagonal Shapes}

Another restricted class of linear constraints was introduced
in~\cite{BalasundaramK89}.  These are of the form
$a v_i + b v_j \leq d$, where $a, b \in \{-1, 0, +1\}$ and $d$
belongs to some computable number family.
Systems of such constraints were called \emph{simple sections}
in~\cite{BalasundaramK89} and have been given the structure of
an abstract domain by A.~Min\'e \cite{Mine01b}.
The resulting \emph{octagon abstract domain} has, due to its syntactic
nature, the same problems outlined in the previous section.
This is why, as explained in detail in \cite{BagnaraHMZ05},
the Parma Polyhedra Library offers a semantic domain of
\emph{octagonal shapes}, for which it provides a widening operator.
This is implemented by the class template
\verb+Octagonal_Shape<T>+, where the class template type parameter \verb+T+
can be instantiated as for bounded difference shapes.\footnote{Support for
octagonal shapes has not yet been included into a formal release.  It is
however complete and available in the PPL's public CVS repository.}
Another feature of this class is that its implementation uses
the strong closure algorithm introduced in \cite{BagnaraHMZ05},
which has lower complexity than the one used in the
Octagon Abstract Domain Library%
\footnote{Until at least version 0.9.8.} \cite{Mine01b,Mine05th}.

\subsection{Grids}

Given $a_1$, \dots,~$a_n$, $b$, $f \in \Zset$,
the \emph{linear congruence relation}
$a_1v_1 + \cdots + a_nv_n \equiv_f b$
denotes the subset of $\Rset^n$
given by
\[
  \biggl\{\,
    \langle q_1, \ldots, q_n \rangle \in \Rset^n
  \bigm|
    \exists \mu \in \Zset \st
      \sum_{i=1}^n a_iq_i = b + \mu f
  \,\biggr\};
\]
when $f \neq 0$, the relation is said to be \emph{proper};
when $f = 0$, the relation is equivalent to (i.e., it denotes the same
hyperplane as) $a_1v_1 + \cdots + a_nv_n = b$.
A \emph{congruence system} is a finite set of congruence relations
and a \emph{grid} is any subset of $\Rset^n$ whose elements
satisfy all the congruences of such a system.
The \emph{grid domain} is the set of all such grids.

\begin{figure}
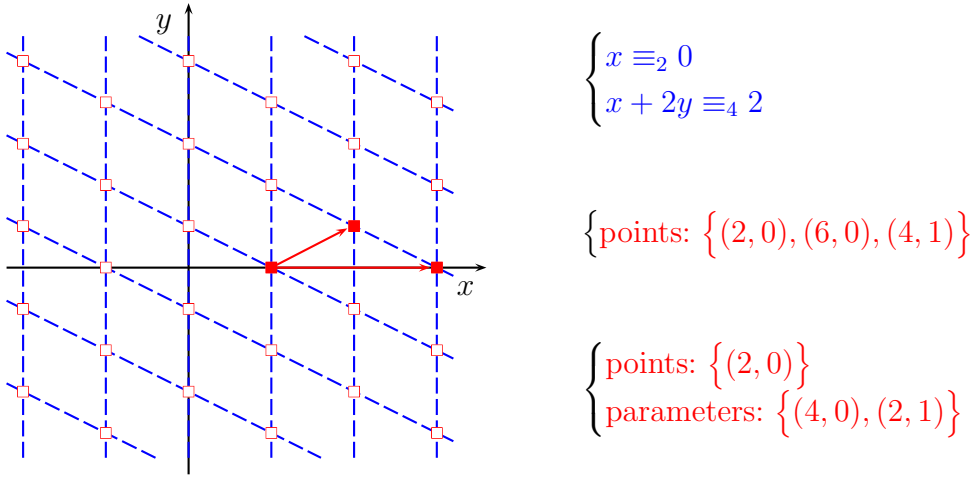

\centering
\begin{minipage}{8cm}
\[
\psset{xunit=1.1cm,yunit=1.1cm,runit=1.1cm}
\psset{origin={1,0.3}}
\pspicture*[](-3.5,0)(3.6,6.2)
\psline{->}(-2.2,3)(3.6,3)
\psline{->}(0,0.5)(0,6.2)
\rput(2.35,2.45){$x$}
\rput(-1.3,5.65){$y$}
\psline[linecolor=blue,linestyle=dashed,dash=6pt 2pt]
(-2,0.7)(-2,5.8)
\psline[linecolor=blue,linestyle=dashed,dash=6pt 2pt]
(-1,0.7)(-1,5.8)
\psline[linecolor=blue,linestyle=dashed,dash=6pt 2pt]
(0,0.7)(0,5.8)
\psline[linecolor=blue,linestyle=dashed,dash=6pt 2pt]
(1,0.7)(1,5.8)
\psline[linecolor=blue,linestyle=dashed,dash=6pt 2pt]
(2,0.7)(2,5.8)
\psline[linecolor=blue,linestyle=dashed,dash=6pt 2pt]
(3,0.7)(3,5.8)
\psline[linecolor=blue,linestyle=dashed,dash=6pt 2pt]
(1.4,5.8)(3.2,4.9)
\psline[linecolor=blue,linestyle=dashed,dash=6pt 2pt]
(-0.6,5.8)(3.2,3.9)
\psline[linecolor=blue,linestyle=dashed,dash=6pt 2pt]
(-2.2,5.6)(3.2,2.9)
\psline[linecolor=blue,linestyle=dashed,dash=6pt 2pt]
(-2.2,4.6)(3.2,1.9)
\psline[linecolor=blue,linestyle=dashed,dash=6pt 2pt]
(-2.2,3.6)(3.2,0.9)
\psline[linecolor=blue,linestyle=dashed,dash=6pt 2pt]
(-2.2,2.6)(1.6,0.7)
\psline[linecolor=blue,linestyle=dashed,dash=6pt 2pt]
(-2.2,1.6)(-0.4,0.7)
\psdots*[linecolor=red,dotstyle=square,dotsize=0.15 0.15]
(-2,1.5)(-2,2.5)(-2,3.5)(-2,4.5)(-2,5.5)
(-1,1)(-1,2)(-1,3)(-1,4)(-1,5)
(0,1.5)(0,2.5)(0,3.5)(0,4.5)(0,5.5)
(1,1)(1,2)(1,3)(1,4)(1,5)
(2,1.5)(2,2.5)(2,3.5)(2,4.5)(2,5.5)
(3,1)(3,2)(3,3)(3,4)(3,5)
\psdots*[linecolor=red,dotstyle=square*,dotsize=0.15 0.15]
(1,3)(3,3)(2,3.5)
\psline[linecolor=red]{->}(1,3)(2.92,3)
\psline[linecolor=red]{->}(1,3)(1.92,3.475)
\endpspicture
\]
\end{minipage}%
\begin{minipage}{5.7cm}
\begin{align*}
&\begin{cases}
  \blue{x \equiv_2 0} \\
  \blue{x + 2y \equiv_4 2}
\end{cases} \\[1cm]
&\begin{cases}
  \red{\text{points: } \bigl\{ (2, 0), (6, 0), (4, 1) \bigr\}}
\end{cases} \\[1cm]
&\begin{cases}
  \red{\text{points: } \bigl\{ (2, 0) \bigr\}} \\
  \red{\text{parameters: } \bigl\{ (4, 0), (2, 1) \bigr\}} \\
\end{cases}
\end{align*}
\end{minipage}
\caption{The double description method for grids}
\label{fig:grid-double-description}
\end{figure}

An example of grid is given in Figure~\ref{fig:grid-double-description},
where the solutions of the congruence relations are given by the dashed
lines and the grid elements by the (filled and unfilled) squares.
The figure shows also that there is an alternative way of describing
the same grid: if we call the points marked by the filled squares
$\mathbf{p}_1 = \left( \begin{smallmatrix}2 \\0\end{smallmatrix} \right)$,
$\mathbf{p}_2 = \left( \begin{smallmatrix}6 \\0\end{smallmatrix} \right)$
and
$\mathbf{p}_3 = \left( \begin{smallmatrix}4 \\1\end{smallmatrix} \right)$,
we can see that the grid is given by
\begin{equation}
\label{eq:affine-integral-combination}
  \{\,
    \pi_1 \mathbf{p}_1 + \pi_2 \mathbf{p}_2 + \pi_3 \mathbf{p}_3 \in \Rset^2
  \mid
    \pi_1, \pi_2, \pi_3 \in \Zset, \pi_1 + \pi_2 + \pi_3 = 1
  \,\}.
\end{equation}
We say that the set of \emph{points}
$\{ \mathbf{p}_1, \mathbf{p}_2, \mathbf{p}_3 \}$ \emph{generates}
the grid.
Some of these generating points can be replaced by \emph{parameters}
that give the direction and spacing for the neighboring points.
Specifically, by subtracting the point $\mathbf{p}_1$ from each of the
other two generating points $\mathbf{p}_2$ and $\mathbf{p}_3$, we obtain
the parameters
$\mathbf{q}_2 = \left( \begin{smallmatrix}4 \\0\end{smallmatrix} \right)$
and
$\mathbf{q}_3 = \left( \begin{smallmatrix}2 \\1\end{smallmatrix} \right)$
and can now express the grid as
\[
  \{\,
    \mathbf{p}_1 + \pi_2 \mathbf{q}_2 + \pi_3 \mathbf{q}_3 \in \Rset^2
  \mid
    \pi_2, \pi_3 \in \Zset
  \,\}.
\]
Notice that, in the generator representation for grids, points and parameters
can have rational, non integer coordinates.

The domain of grids, as briefly described above, has been introduced
by P.~Granger \cite{Granger91,Granger97} and its design has been
completed in \cite{BagnaraDHMZ06a,BagnaraDHMZ06b}
by including refined algorithms and new operators
for affine images, preimages and their generalizations, grid-difference
and widening.
The Parma Polyhedra Library includes the first truly complete
implementation of this abstract domain by means of the \verb+Grid+ class.
Congruence relations and systems thereof are realized by the classes
\verb+Congruence+ and \verb+Congruence_System+, and likewise
for generators with the classes \verb+Grid_Generator+,
and \verb+Grid_Generator_System+.

A more restricted domain, the domain of \emph{integral lattices}
has been partly implemented in PolyLib~\cite{Loechner99}
following the approach in~\cite{QuintonRR96,QuintonRR97}.
An integral lattice in dimension $n$ is the grid generated by
the affine integral combination
---as in \eqref{eq:affine-integral-combination}---
of exactly $n+1$ affinely independent points that are additionally
bound to have integer coordinates.
These restrictions have the consequences that only the generator representation
is supported and, as explained in \cite{BagnaraDHMZ06a,BagnaraDHMZ06b},
the reduced expressivity
has an impact on the possibility to solve concrete analysis problems.
The implementation of the domain of integral lattices in PolyLib is
incomplete in the sense that it misses domain operators such as
the join. Instead, it provides a union operator
which, given two lattices returns a \emph{set} of lattices
representing the precise set union of the points of the given lattices.
Similarly the difference operation on a pair of lattices
returns a set of lattices whose union contains the exact
set difference of their points.
Moreover, although image and preimage operators are supported
by PolyLib, as the integral lattice must be full dimensional,
only invertible image operators are allowed.

\subsection{Powersets}

For applications requiring a high degree of precision, the Parma Polyhedra
Library provides a generic implementation of the \emph{finite powerset}
construction \cite{Bagnara98SCP,BagnaraHZ06STTT}.
This upgrades an abstract domain into a refined one where finite
disjunctions of non redundant elements are precisely representable.
The construction is implemented generically by the class template
\verb+Powerset<D>+, where the type parameter \verb+D+ must provide
about ten methods and operators implementing, among other things,
an entailment predicate and operators for obtaining an upper bound
and the ``meet'' of two given elements.  No other requirements are
imposed and, in fact, the test suite of the PPL includes a program
where \verb+Powerset<D>+ is instantiated with a non numerical domain.

The class template \verb+Pointset_Powerset<PS>+ provides a specialization
of class \verb+Powerset<D>+ that is suitable for the instantiation with the
``semantic'' numerical domains of the PPL: (C or NNC) polyhedra, bounded
difference shapes, octagonal shapes, grids
and combinations thereof.
A most notable and, at the time of writing, unique feature of this
implementation is the provision ---in addition to the extrapolation
operator proposed in \cite{BultanGP99}--- of provably correct,
\emph{certificate-based} widening operators that lift the widening operators
defined on the underlying domain \verb+PS+ \cite{BagnaraHZ06STTT}.

\subsection{Linear Programming Problems}
\label{sec:linear-programming-problems}

The library includes a Linear Programming (LP) solver based
on the simplex algorithm and using exact arithmetic.
Note that the absence of rounding errors is essential for most
(if not all) of the intended applications.
Since it is common practice to solve an LP problem
by exploiting duality results, correctness may be lost even if
\emph{controlled rounding} is used; this is because a feasible but possibly
non-optimal solution for the dual of an LP problem may actually correspond
to an unfeasible solution for the original LP problem.

The LP solver interface allows for both satisfiability checks
and optimization of linear objective functions.
A limited form of incrementality allows for the efficient re-optimization
of an LP problem after modification of the objective function.
Ongoing implementation work is focusing on improving the
efficiency of the solver as well as providing better support
for incremental computations, so as to also allow for the efficient
re-optimization of the LP problem after a modification of the
feasible region caused by the addition of further constraints.%
\footnote{The incremental LP solver has not yet been included into
a formal release.  It is however complete and available in
the PPL's public CVS repository.}

\section{Main Features}
\label{sec:features}

In this section we will briefly review the main features of the Parma
Polyhedra Library.  We will focus on usability, on the absence of
arbitrary limits due to the use of fully dynamic data structures,
on some of the measures that contribute to the library's robustness,
on the support for resource-bounded computations, on the possibility
to use machine integer coefficients without compromising correctness,
on portability and on the availability of complete documentation.

\subsection{Usability}

By ``usability'' of the Parma Polyhedra Library we actually mean two things:
\begin{enumerate}
\item
that the library provides natural, easy to use interfaces that can be used,
even by the non expert, to quickly prototype an application;
\item
that, nonetheless, the library and its interfaces provide all the
functionalities that allow their use in the development of professional,
reliable applications.
\end{enumerate}
In other words, simplicity of the interfaces has not been obtained
with simplistic solutions or by the omission of functionalities.

As mentioned before particular care has been taken (to the point of
developing the necessary theoretical concepts) in the complete
decoupling of the user interfaces from all implementation details.
So, the internal representation of, say, constraints, congruences,
generators and systems thereof need not concern the client application.
All the user interfaces, whatever language they interface to,
refer to high-level concepts
and never to their possible implementation in terms of vectors, matrices
or other data structures.  For instance, unlike PolyLib and
New Polka, implementation devices (such as so-called
\emph{positivity constraints} \cite{Wilde93th}
or \emph{$\epsilon$-representations} \cite{BagnaraHZ05FAC,HalbwachsPR94})
never surface at the user level and need not concern the client
application.  As another example, a user of the Octagon Abstract Domain
Library must know that octagons are represented there by means of
difference-bound matrices, that some of the operations on octagons
do ``close'' these matrices, and that one argument to the widening
is better closed for improved accuracy while the other should \emph{not}
be closed as this would prevent convergence.

The Parma Polyhedra Library currently offers, besides the \Cplusplus{}
interface, a portable Prolog interface and a C interface.
The Prolog interface is ``portable'' in that it supports (despite the
lack of standardization of Prolog foreign language interfaces) six
major Prolog systems: Ciao, GNU Prolog, SWI-Prolog, SICStus, XSB and
YAP.
The C interface is particularly important, as it allows to interface
the PPL with almost anything else: for example, third parties have
already built interfaces for Haskell, Java and Objective Caml.
The design of the interfaces to directly supported languages has been
focused on ensuring that programmers can use the library following the most
natural programming style in \emph{that} language.
As a simple example, in the appropriate contexts, `\texttt{X~<~5*Z}'
and `\texttt{X~+~2*Y~+~5*Z~>=~7}' is valid syntax expressing
a strict and a non-strict inequality, both in the \Cplusplus{}
and the Prolog interfaces.  This can be done because both languages
allow to override (or freely interpret) operators and provide
exceptions as a powerful method of reporting run-time errors.
Here is how a NNC polyhedron in a space of dimension~$3$ can be created
using the \Cplusplus{} interface:
\bigskip
\lstset{language=C++}
{\small
\begin{lstlisting}[frame=lines]

#include <ppl.hh>

namespace PPL = Parma_Polyhedra_Library;

  ...
  PPL::Variable X(0);
  PPL::Variable Y(1);
  PPL::Variable Z(2);
  PPL::NNC_Polyhedron ph(3, PPL::UNIVERSE);
  ph.add_constraint(X + 2*Y + 5*Z >= 7);
  ph.add_constraint(X < 5*Z);
  ...
\end{lstlisting}
} 
And here is how the same polyhedron can be created
using the Prolog interface:
\newpage
\lstset{language=Prolog}
{\small
\begin{lstlisting}[frame=lines]

  ...
  numbervars([X, Y, Z], 0, _),
  ppl_new_NNC_Polyhedron_from_constraints(
    [X + 2*Y + 5*Z >= 7, X < 5*Z],
    PH
  ),
  ...
\end{lstlisting}
} 
In standard C things are more complicate as the language syntax does not
allow to represent, say, constraints as easily.\footnote{Unless one
represents them with strings.  While a string-based interface is
certainly possible, we do not believe it would make things simpler:
strings would have to be parsed at run-time and parsing errors would
have to be properly handled.}
Thus, in order to build constraints the C application will have to build
the linear expressions occurring in them and the memory used to hold these
intermediate data structures will have to be explicitly managed,
unless a conservative garbage collector is used.
Moreover, lack of exceptions means that non-trivial error detection
and handling will demand significantly more effort in C than in \Cplusplus{}
or Prolog (\emph{all} the functions of the C interface return an \verb+int+:
a negative value indicates an error has occurred).
The best approach to development using the C interface is to begin
by developing a layer of interface functions that are suited to
the application at hand:
the PPL's C interface provides all the required services.

Still on the subject of ease of use, the numeric abstractions provided
by the library provide similar interfaces. A high degree of
integration is obtained through the adoption of common data types and
the availability of methods specifically designed for
interoperability.
As a consequence, the specification and implementation of
new abstractions and tools can be easily obtained by composition
of the available services.
As a simple example, the code in Listing~\ref{lst:lp_solver} is
exploiting the LP solver capabilities to efficiently compute (and
print) all the upper bounds for the variables corresponding to the
dimensions of a closed polyhedron.

\lstset{language=[ISO]C++}
{\small
\begin{lstlisting}[caption={Printing upper bounds for all variables using the LP solver},label=lst:lp_solver,frame=lines,float]

#include <ppl.hh>
#include <iostream>

using namespace Parma_Polyhedra_Library;
using namespace Parma_Polyhedra_Library::IO_Operators;

void print_upper_bounds(const C_Polyhedron& ph) {
  LP_Problem lp(ph.constraints());
  // Check the satisfiability of the problem.
  if (!lp.is_satisfiable()) {
    std::cout << "unsatisfiable" << std::endl;
    return;
  }
  // Print the upper bound of each variable.
  lp.set_optimization_mode(MAXIMIZATION);
  Generator g(point());
  Coefficient num, den;
  const dimension_type dim = ph.space_dimension();
  for (dimension_type i = 0; i < dim; ++i) {
    Variable x(i);
    lp.set_objective_function(x);
    LP_Problem_Status status = lp.solve();
    if (status == UNBOUNDED_LP_PROBLEM)
      std::cout << x << " < +infty" << std::endl;
    else {
      assert(status == OPTIMIZED_LP_PROBLEM);
      g = lp.optimizing_point();
      lp.evaluate_objective_function(g, num, den);
      std::cout << x << " <= " << num << "/" << den
                << std::endl;
    }
  }
}
\end{lstlisting}
}

Note that the code in Listing~\ref{lst:lp_solver} is taking advantage
of a limited form of incrementality provided by the LP solver: the
check for satisfiability (corresponding to the first phase of the
simplex algorithm) is executed only once and it is not repeated when
optimizing the different objective functions (namely, only the second
phase of the simplex algorithm is executed in the for-loop).  Code
similar to the one above is actually used in the library itself to
precisely approximate a polyhedron by means of a bounded difference or
octagonal shape without incurring the potentially high cost of
converting from the constraint to the generator representation of the
polyhedron.

\subsection{Absence of Arbitrary Limits}

The only real restrictions imposed by the library on the client
application are those caused by limitations of the available virtual
memory.
All data structures are fully dynamic and automatically expand
(in amortized constant time) and shrink in a way that is completely
transparent to the user, ensuring the best use of available memory.

In contrast, in the PolyLib, New Polka and HyTech
libraries, matrices of coefficients, which are the main data
structures used to represent polyhedra, cannot grow dynamically
and the client application is ultimately responsible for specifying
their dimensions.
Since the worst case space complexity of the methods employed is
exponential, in general the client application cannot make a safe and
practical choice:
specifying small dimensions may provoke a run-time failure;
generous dimensions may waste significant amounts of memory
and, again, result in unnecessary run-time failures.

\subsection{Robustness}

The clean separation between interface and implementation, among
other important advantages, allows for the adoption of incremental
and lazy computation techniques.
The increased efficiency due to these techniques amply repays the
cost of the interface checks that contribute to the library's
robustness, which, as it will be explained in the sequel,
is one of its most important features.

First, the library systematically checks all the interface invariants
and throws an exception if any one of them is violated.
This makes it very difficult to inadvertently create invalid objects
and greatly simplifies the debugging of client applications.
Secondly, the library is exception-safe, that is, it never leaks
resources or leaves invalid object fragments around, even in the
presence of exceptions.
In particular, if an exception is raised, then all the memory
allocated by the failed computation is discarded.
These features allows applications using the PPL to use timeouts
or to sensibly deal with run-time errors such as arithmetic overflows
and out-of-memory conditions, so as to continue the computation in
a reliable state (see below for more on this subject).
It is important to stress that, while error handling and recovery are
somewhat optional features (there may be no interest in continuing the
computation after an error has occurred), error detection should be
considered a mandatory feature in the field of system's analysis and
verification, since failure to detect an error can easily result into
undefined (i.e., completely unpredictable) behavior,
therefore compromising any statement about correctness.

For comparison, PolyLib detects only some errors, sometimes
setting a flag and sometimes printing a message and aborting, whereas
New Polka and the HyTech libraries detect some errors, print
an error message and abort.  The Octagon Abstract Domain Library makes
no attempt at detecting errors.  None of these libraries perform a
systematic check of interface invariants.

\subsection{Resource-Bounded Computations}

The library, thanks to its exception-safety characteristics, naturally
supports resource-bounded computations, that is, computations that are
limited in the amount of CPU time or virtual memory or both.  The
client application can take advantage of this feature by attempting a
computation that is potentially very expensive imposing a maximum
limit on the resources to be used.  Should this bound be exceeded, an
exception is raised and the client application can resort to a
simplified computation (possibly using a simpler numerical
abstraction) trusting that the PPL will release all the resources
allocated by the interrupted computation.  With these facilities at
hand, users of the library can quite easily code resource-bounded or,
at least, resource-conscious numerical abstractions.
An example is shown in Listing~\ref{lst:decl-rcc}.
\lstset{language=[ISO]C++}
{\small
\begin{lstlisting}[caption={Declaration of a class for resource-conscious computations},label=lst:decl-rcc,frame=lines,float]

#include <ppl.hh>
#include <pwl.hh>
#include <stdexcept>

namespace PPL = Parma_Polyhedra_Library;
namespace PWL = Parma_Watchdog_Library;

class Up_Appr_Polyhedron {
private:
  // The type of the main polyhedral abstraction.
  typedef PPL::C_Polyhedron PH;

  // The type of a simpler polyhedral abstraction.
  typedef PPL::BD_Shape<mpq_class> SPH;

  PH ph;

  // Timeout in hundredth of a second; 0 for no timeout.
  static unsigned long timeout_hs;

  class Timeout : public PPL::Throwable {
    ...
  };

  static Timeout Timeout_object;

public:
  static void set_timeout(unsigned long n);
  ...
  void intersection_assign(const Up_Appr_Polyhedron& y);
  ...
};
\end{lstlisting}
}
This uses the \emph{Parma Watchdog Library} (PWL), a library that virtualizes
the interval timers for any XSI-conforming implementation of UNIX.%
\footnote{XSI it is the core application
programming interface for C and shell programming for systems conforming
to the \emph{Single UNIX Specification}.}
The PWL, which is currently distributed with the PPL, gives an
application the ability to work with an unbounded number of independent
``watchdog'' timers.

The example class \verb+Up_Appr_Polyhedron+ is meant to provide
convex polyhedra with ``upward approximated'', resource-conscious
operations.  As the main representation it uses closed convex polyhedra;
the user can set a timeout for the operations and externally impose a
limit on the virtual memory available to the process.
When a resource limit is reached, the class temporarily switches
to a simpler family of polyhedra: bounded difference shapes.
Listing~\ref{lst:def-rcc} shows a simple implementation for the intersection
operation.
\lstset{language=[ISO]C++}
{\small
\begin{lstlisting}[caption={Definition of a resource-conscious intersection method},label=lst:def-rcc,frame=lines,float]

void Up_Appr_Polyhedron
::intersection_assign(const Up_Appr_Polyhedron& y) {
  if (timeout_hs == 0) {
    // No timeout: do the operation directly.
    ph.intersection_assign(y.ph);
    return;
  }

  // Save copies of `ph' and `y.ph': they may be needed
  // to recover from a resources exhaustion condition.
  PH xph_copy = ph;
  PH yph_copy = y.ph;

  try {
    PWL::Watchdog w(timeout_hs,
		    PPL::abandon_expensive_computations,
		    Timeout_object);
    ph.intersection_assign(y.ph);
    PPL::abandon_expensive_computations = 0;
    return;
  }
  catch (const Timeout&) {
    // Timeout expired.
  }
  catch (const std::bad_alloc&) {
    // Out of memory.
  }

  // Resources exhausted: use simpler polyhedra.
  PPL::abandon_expensive_computations = 0;

  SPH xph_simple(xph_copy, PPL::POLYNOMIAL_COMPLEXITY);
  SPH yph_simple(yph_copy, PPL::POLYNOMIAL_COMPLEXITY);
  xph_simple.intersection_assign(yph_simple);
  ph = PH(xph_simple.minimized_constraints());

  // Restore `y.ph'.
  const_cast<PH&>(y.ph) = yph_copy;
}
\end{lstlisting}
}
If no timeout has been requested, then the operation is performed
without any overhead.  Otherwise the polyhedra are copied, a watchdog
timer, \verb+w+, is set and the operation is attempted.
When \verb+w+ expires, the PPL is asked to abandon all expensive
computations and to throw \verb+Timeout_object+.  Alternatively,
if the process exceeds the virtual memory it has been allotted,
then the \verb+bad_alloc+ standard exception will be thrown.
If none of those happen, then control will be returned to the caller.
Otherwise, bounded difference shapes approximating the argument
polyhedra will be computed using a polynomial complexity
method,%
\footnote{As mentioned in Section~\ref{sec:linear-programming-problems},
it is also possible to compute these approximations using a
``simplex complexity'' algorithm (i.e., theoretically exponential but
very efficient in practice).}
these shapes will be intersected and the intersection will be used
to construct the resulting convex polyhedron.

The technique illustrated in a simplified way by Listings~\ref{lst:decl-rcc}
and~\ref{lst:def-rcc} is quite powerful and allows to
deal with the complexity-precision trade-off in a very flexible way.
In particular, it is possible, thanks to the PWL, to work with
multiple timers: while individual polyhedra operations can be guarded
by a timer, other timers can monitor operations of greater
granularity, such as entire analysis phases.  When an analysis phase
is taking too much, then the timeouts used for the individual
operations can be shortened or the analyzer can switch to a totally
different, less complex analysis technique.  It is worth observing
that, notwithstanding the friendliness of the PPL's user interfaces,
professional applications in the field of system's analysis and
verification are not expected to be directly based on the abstractions
provided by the library.  Rather, the PPL abstractions have been designed
so as to serve as building blocks for the actual analysis domains:
in this field the complexity-precision trade-off is often so serious
that the right way to face it is, by necessity, application-dependent.

\subsection{Unbounded or Native Integer Coefficients}
\label{sec:unbounded-or-native-integer-coefficients}

For the representation of general convex polyhedra, with the default
configuration, the Parma Polyhedra Library uses unbounded precision integers.
On the other hand, if speed is important and the numerical coefficients
involved are likely to be small, applications may use native integers
(8, 16, 32 and 64 bit integers are supported by the PPL).
This is a safe strategy since, when using native integers, the library
also performs systematic (yet efficient) overflow detection.
It is thus possible to adopt an approach whereby computations are first
attempted with native integers.  If a computation runs to completion,
the user can be certain that no overflow occurred.  Otherwise an exception
is raised (as in the case seen before for resource-bounded computations),
so that the client application can be restarted with bigger native
integers or with unbounded integers.
This is another application of the library's exception-safety, as one
can rather simply code the above approach as follows:
\bigskip
\lstset{language=[ISO]C++}
{\small
\begin{lstlisting}[frame=lines]

  try {
    // Analyze with 64-bit coefficients.
    ...
  }
  catch (const std::overflow_error&) {
    // Analyze with unbounded coefficients.
    ...
  }
  ...
\end{lstlisting}
} 
Again, the client application does not need to be concerned about
the resources allocated by the PPL during the computation of the
\verb+try+ block: everything will be deallocated automatically.

Concerning other libraries, PolyLib and New Polka can use unbounded
integers as coefficients, whereas the library of HyTech does
not support them.  Differently from the PPL, these libraries use
finite integral types without any mechanism for overflow detection.
Technically speaking and according to the C standard (the language in
which they are written), this means that the effects of an overflow
are completely undefined, i.e., client applications cannot make
any assumption about what can happen should an overflow occur.
In addition, PolyLib (and, according to \cite{BerardF99},
some versions of HyTech) can use floating point values,
in which case underflows and rounding errors, in addition to overflows,
can affect the results.

\subsection{Portability and Documentation}

Great care has been taken to ensure the portability of the PPL.
The library is written in standard \Cplusplus{}, it follows all
the available applicable standards and uses sophisticated
automatic configuration mechanisms.  It is known to run on all
major UNIX-like operating systems, on Mac~OS~X (whether Intel-
or PowerPC-based) and on Windows (via Cygwin or MinGW).

A big investment has also been made on documentation and at several
levels.  First, the theoretical underpinnings have been thoroughly
investigated, integrated when necessary and written down: an extensive
bibliography is available on the PPL web site.  Secondly, during the
entire development of the library, the quality, accessibility and
completeness of the documentation has always been given a particular
emphasis: while some parts of the library need more work in this respect,
the vast majority of the code is thoroughly documented and some parts
of it approach the ideal of ``literate programming.''

The library has been documented using the Doxygen
tool.\footnote{\url{http://www.doxygen.org}.}
Doxygen is a documentation system for C++, C, Java, and other languages
that allows to generate high-quality documentation from a collection
of documented source files.  The source files can be documented by means
of ordinary comments, that can be placed near the program elements being
documented: just above the declaration or definition of a member, class
or namespace, for instance.  This makes it much easier to keep the
documentation consistent with the actual source code.  Moreover, Doxygen
allows the typesetting of mathematical formulas within comments by means
of the relevant subset of \LaTeX{}, which is an important feature for
a project like the PPL.  It is also able to automatically
extract the code structure and use this information to generate
include dependency graphs, inheritance diagrams, and collaboration diagrams.
Doxygen can generate documentation in various formats, such as HTML,
PostScript and PDF.  The HTML and PDF output are fully hyperlinked,
a feature that greatly facilitates ``navigation'' in the available
documentation.

The Parma Polyhedra Library is equipped with two manuals generated
with the help of Doxygen: a user's manual, containing all and only the
information needed by people wishing to use the library
\cite{PPL-USER-0-9}; and a developer's reference manual that contains,
in addition, all the details concerning the library implementation
\cite{PPL-DEVREF-0-9}.
All manuals are available, in various formats, from the PPL web site
and the user's manual is also included in each source distribution.

\section{Efficiency}
\label{sec:efficiency}

One natural question is how does the efficiency of the
Parma Polyhedra Library compare with that of other polyhedra libraries.
Of course, such a question does not have a definite answer.  Apart
from clarifying whether CPU or memory efficiency or both are the
intended measures of interest, the answer will depend on the targeted
applications: with different applications the results can vary wildly.
Moreover, even within the same application, big variations may be
observed for different inputs.
For these reasons, it must be admitted that the only way to meaningfully
assess the performance of the library is with respect to a particular
application, a particular set of problem instances, and a particular
definition of `performance'.

For the same reasons, it is nonetheless instructive to compare the
performance of various polyhedra libraries on a well-defined problem
with a large set of freely available inputs.
One such problem, called \emph{vertex/facet enumeration}, is particularly
relevant for implementations based on the Double Description method
such as the Parma Polyhedra Library, New Polka and PolyLib,
as this problem has to be solved whenever one description has to be
converted into the other one.
The vertex/facet enumeration problem is a well-studied one and
several systems have been expressly developed to solve it.
We have thus compared the above mentioned libraries with
the following (in parentheses, the versions we have tested):
\begin{itemize}
\item
cddlib (version 0.94b), a C implementation of the Double Description
method, by K.~Fukuda
\cite{FukudaP96};%
\footnote{\url{http://www.cs.mcgill.ca/~fukuda/soft/cdd_home/cdd.html}.}
\item
lrslib (version 0.42b), a C implementation of the reverse search algorithm
for vertex enumeration/convex hull problems, by D.~Avis
\cite{Avis98,Avis00};\footnote{\url{http://cgm.cs.mcgill.ca/~avis/C/lrs.html}.}
\item
pd (version 1.7), a C program implementing a primal-dual algorithm
using rational arithmetic, by A.~Marzetta and maintained by D.~Bremner
\cite{BremnerFM98}.%
\footnote{\url{http://www.cs.unb.ca/profs/bremner/pd/}.}
\end{itemize}
Both cddlib and lrslib come with driver programs that support a polyhedra
input format that was introduced by K.~Fukuda and extended by D.~Avis;
this input format is also supported by the pd program.
The distributions of cddlib and lrslib provide more than 100 different inputs
of varying complexity for these programs.
Driver programs that can read the same input format and use the PPL,
New Polka and PolyLib are part of the PPL distribution since version 0.7.

The tests have been performed on a PC equipped with an AMD Athlon 2800+
with 1 GB of RAM and running GNU/Linux and GMP version 4.2.
All the software has been compiled with GCC 4.0.3 at the optimization level
that is the default for each package (i.e., the PPL was compiled with
`\verb/-O2/', its default; PolyLib, cddlib, and pd with \verb/-O2/;
New Polka and lrslib with `\verb/-O3/').
The obtained running times, in seconds, are reported
in Tables~\ref{tab:efficiency-vertex-enumeration}
and~\ref{tab:efficiency-vertex-enumeration-hard}.%
\footnote{Filenames have been shortened to fit the table on the page:
in particular the \texttt{.ext} and \texttt{.ine} extensions have been
shortened to \texttt{.e} and \texttt{.i}, respectively; moreover, the file
called \texttt{integralpoints.ine} has been renamed \texttt{integpoints.i}.}
The entries marked with `n.a.' in the pd's column indicate the problems
that cannot be solved by pd, which can only handle polyhedra that contain
the origin.  Entries marked with `ovfl' indicate the problems on which
pd runs into an arithmetic overflow.
It should be noted that strictly speaking, lrslib solves a slightly
different, easier problem than the one solved by the other systems:
while the latter guarantee the result is minimized, the output of
lrslib may contain duplicate rays.

{
\renewcommand{\arraystretch}{0.91}
\small
\begin{longtable}{l......}
\caption{Efficiency on vertex enumeration%
\label{tab:efficiency-vertex-enumeration}} \\
\hline
\multicolumn{1}{c}{input} &
\multicolumn{1}{c}{PPL} &
\multicolumn{1}{c}{New Polka} &
\multicolumn{1}{c}{PolyLib} &
\multicolumn{1}{c}{cddlib} &
\multicolumn{1}{c}{lrslib} &
\multicolumn{1}{c}{pd} \\
\hline
\endfirsthead
\caption[]{Efficiency on vertex enumeration (continued)} \\
\hline
\multicolumn{1}{c}{input} &
\multicolumn{1}{c}{PPL} &
\multicolumn{1}{c}{New Polka} &
\multicolumn{1}{c}{PolyLib} &
\multicolumn{1}{c}{cddlib} &
\multicolumn{1}{c}{lrslib} &
\multicolumn{1}{c}{pd} \\
\hline
\endhead
\hline
\endfoot
\verb/ccc4.e/&0.00&0.11&0.03&0.00&0.00&\multicolumn{1}{c}{n.a.}\\
\verb/ccc5.e/&0.00&0.10&0.04&0.02&0.00&\multicolumn{1}{c}{n.a.}\\
\verb/ccc6.e/&0.03&0.14&0.26&0.61&3.12&\multicolumn{1}{c}{n.a.}\\
\verb/ccp4.e/&0.00&0.10&0.02&0.00&0.00&0.01 \\
\verb/ccp5.e/&0.00&0.10&0.04&0.02&0.00&5.36 \\
\verb/ccp6.e/&0.05&0.15&0.31&0.90&3.95&\multicolumn{1}{c}{$>1h$} \\
\verb/cp4.e/&0.00&0.08&0.03&0.00&0.00&0.01 \\
\verb/cp5.e/&0.00&0.12&0.05&0.02&0.00&5.29 \\
\verb/cp6.e/&0.05&0.18&0.30&0.88&3.86&\multicolumn{1}{c}{$>1h$} \\
\verb/cube.e/&0.00&0.05&0.02&0.00&0.00&0.00 \\
\verb/cut16_11.e/&0.00&0.12&0.04&0.02&0.00&3.86 \\
\verb/cut32_16.e/&0.05&0.17&0.28&0.91&4.32&\multicolumn{1}{c}{$>1h$} \\
\verb/cyclic10-4.e/&0.00&0.07&0.02&0.00&0.00&0.00 \\
\verb/cyclic12-6.e/&0.00&0.08&0.03&0.01&0.00&0.44 \\
\verb/cyclic14-8.e/&0.00&0.09&0.04&0.06&0.01&172.10 \\
\verb/cyclic16-10.e/&0.02&0.14&0.07&0.24&0.04&\multicolumn{1}{c}{$>1h$} \\
\verb/dcube10.e/&0.02&0.13&0.10&0.23&0.02&\multicolumn{1}{c}{$>1h$} \\
\verb/dcube12.e/&0.17&0.22&2.66&1.29&0.10&\multicolumn{1}{c}{$>1h$} \\
\verb/dcube3.e/&0.00&0.08&0.02&0.00&0.00&0.00 \\
\verb/dcube6.e/&0.00&0.07&0.03&0.00&0.00&0.35 \\
\verb/dcube8.e/&0.00&0.12&0.03&0.04&0.00&119.53 \\
\verb/irbox20-4.e/&0.00&0.08&0.01&0.01&0.00&0.01 \\
\verb/irbox200-4.e/&0.02&0.07&0.02&0.98&0.05&0.18 \\
\verb/mp5.e/&0.00&0.11&0.05&0.59&0.84&3.93 \\
\verb/prodst62.e/&34.78&161.09&123.28&\multicolumn{1}{c}{$>1h$}&\multicolumn{1}{c}{$>1h$}&\multicolumn{1}{c}{ovfl} \\
\verb/redcheck.e/&0.00&0.06&0.01&0.00&0.00&0.00 \\
\verb/reg24-5.e/&0.00&0.08&0.03&0.01&0.00&0.01 \\
\verb/reg600-5_m.e/&5.05&19.12&12.37&135.49&33.63&\multicolumn{1}{c}{n.a.}\\
\verb/samplev1.e/&0.00&0.09&0.02&0.00&0.00&\multicolumn{1}{c}{n.a.}\\
\verb/samplev2.e/&0.00&0.06&0.02&0.00&0.00&\multicolumn{1}{c}{n.a.}\\
\verb/samplev3.e/&0.00&0.07&0.02&0.00&0.00&\multicolumn{1}{c}{n.a.}\\
\verb/tsp5.e/&0.00&0.12&0.04&0.00&0.00&\multicolumn{1}{c}{n.a.}\\
\verb/allzero.i/&0.00&0.06&0.01&0.00&0.00&\multicolumn{1}{c}{n.a.}\\
\verb/cp4.i/&0.00&0.09&0.02&0.00&0.00&\multicolumn{1}{c}{n.a.}\\
\verb/cp5.i/&0.01&0.12&0.05&0.14&5.91&\multicolumn{1}{c}{n.a.}\\
\verb/cross10.i/&0.08&0.22&0.14&10.50&\multicolumn{1}{c}{$>1h$}&1.38 \\
\verb/cross12.i/&0.84&3.04&2.86&166.15&\multicolumn{1}{c}{$>1h$}&15.24 \\
\verb/cross4.i/&0.00&0.07&0.02&0.00&0.00&0.00 \\
\verb/cross6.i/&0.00&0.06&0.03&0.04&0.09&0.01 \\
\verb/cross8.i/&0.01&0.07&0.03&0.54&28.90&0.14 \\
\verb/cube.i/&0.00&0.09&0.02&0.00&0.00&0.00 \\
\verb/cube10.i/&0.02&0.15&0.16&0.24&0.03&\multicolumn{1}{c}{$>1h$} \\
\verb/cube12.i/&0.19&0.32&3.70&1.35&0.18&\multicolumn{1}{c}{$>1h$} \\
\verb/cube3.i/&0.00&0.06&0.02&0.00&0.00&0.00 \\
\verb/cube6.i/&0.00&0.10&0.03&0.01&0.00&0.34 \\
\verb/cube8.i/&0.00&0.10&0.05&0.06&0.00&123.23 \\
\verb/cubetop.i/&0.00&0.08&0.02&0.00&0.00&\multicolumn{1}{c}{n.a.}\\
\verb/cubocta.i/&0.00&0.10&0.02&0.00&0.00&0.00 \\
\verb/cyc.i/&0.00&0.08&0.01&0.00&0.00&0.00 \\
\verb/cyclic17_8.i/&0.04&0.15&0.14&0.32&0.08&\multicolumn{1}{c}{$>1h$} \\
\verb/diamond.i/&0.00&0.08&0.01&0.00&0.00&0.00 \\
\verb/dodeca_m.i/&0.00&0.07&0.02&0.00&0.00&\multicolumn{1}{c}{n.a.}\\
\verb/ex1.i/&0.00&0.07&0.02&0.00&0.00&\multicolumn{1}{c}{n.a.}\\
\verb/grcubocta.i/&0.00&0.06&0.02&0.01&0.00&0.01 \\
\verb/hexocta.i/&0.00&0.05&0.01&0.02&0.00&0.01 \\
\verb/icododeca_m.i/&0.00&0.07&0.02&0.05&0.00&\multicolumn{1}{c}{n.a.}\\
\verb/in0.i/&0.00&0.06&0.02&0.00&0.00&\multicolumn{1}{c}{n.a.}\\
\verb/in1.i/&0.00&0.08&0.02&0.01&0.00&\multicolumn{1}{c}{n.a.}\\
\verb/in2.i/&0.00&0.08&0.03&0.00&0.00&\multicolumn{1}{c}{n.a.}\\
\verb/in3.i/&0.00&0.05&0.02&0.00&0.00&\multicolumn{1}{c}{n.a.}\\
\verb/in4.i/&0.00&0.10&0.03&0.01&0.00&\multicolumn{1}{c}{n.a.}\\
\verb/in5.i/&0.00&0.07&0.04&0.03&0.00&\multicolumn{1}{c}{n.a.}\\
\verb/in6.i/&0.02&0.14&0.09&0.42&0.03&\multicolumn{1}{c}{n.a.}\\
\verb/in7.i/&0.18&0.40&0.44&0.96&0.08&\multicolumn{1}{c}{n.a.}\\
\verb/infeas.i/&0.00&0.09&0.02&0.00&0.00&\multicolumn{1}{c}{n.a.}\\
\verb/integpoints.i/&0.00&0.10&0.04&0.05&0.00&\multicolumn{1}{c}{n.a.}\\
\verb/kkd18_4.i/&0.00&0.10&0.02&0.02&0.00&\multicolumn{1}{c}{n.a.}\\
\verb/kkd27_5.i/&0.05&0.12&0.07&0.08&0.01&\multicolumn{1}{c}{n.a.}\\
\verb/kkd38_6.i/&2.95&5.26&1.52&0.32&0.05&\multicolumn{1}{c}{n.a.}\\
\verb/kq20_11_m.i/&0.19&0.41&0.45&0.99&0.08&\multicolumn{1}{c}{n.a.}\\
\verb/metric40_11.i/&0.00&0.11&0.04&0.07&0.57&\multicolumn{1}{c}{n.a.}\\
\verb/metric80_16.i/&0.13&0.24&0.07&0.54&32.09&\multicolumn{1}{c}{n.a.}\\
\verb/mit31-20.i/&23.34&27.21&115.00&103.19&25.53&\multicolumn{1}{c}{$>1h$} \\
\verb/mp5.i/&0.00&0.08&0.04&0.06&0.58&\multicolumn{1}{c}{n.a.}\\
\verb/mp5a.i/&0.00&0.10&0.03&0.06&0.57&\multicolumn{1}{c}{n.a.}\\
\verb/mp6.i/&0.33&0.42&0.75&4.66&1177.96&\multicolumn{1}{c}{n.a.}\\
\verb/nonfull.i/&0.00&0.06&0.02&0.00&0.00&\multicolumn{1}{c}{n.a.}\\
\verb/origin.i/&0.00&0.08&0.02&0.00&0.00&\multicolumn{1}{c}{n.a.}\\
\verb/project1_m.i/&0.00&0.10&0.04&0.03&0.00&1.10 \\
\verb/project1res.i/&0.00&0.08&0.02&0.00&0.00&0.00 \\
\verb/project2_m.i/&0.02&0.09&0.05&0.46&0.13&\multicolumn{1}{c}{n.a.}\\
\verb/project2res.i/&0.00&0.08&0.02&0.08&0.01&\multicolumn{1}{c}{n.a.}\\
\verb/rcubocta.i/&0.00&0.08&0.01&0.01&0.00&0.01 \\
\verb/reg24-5.i/&0.00&0.08&0.02&0.01&0.00&0.01 \\
\verb/rhomtria_m.i/&0.00&0.09&0.02&0.04&0.00&\multicolumn{1}{c}{n.a.}\\
\verb/sample.i/&0.00&0.08&0.01&0.00&0.00&0.00 \\
\verb/sampleh1.i/&0.00&0.05&0.02&0.00&0.00&\multicolumn{1}{c}{n.a.}\\
\verb/sampleh2.i/&0.00&0.06&0.01&0.00&0.00&\multicolumn{1}{c}{n.a.}\\
\verb/sampleh3.i/&0.00&0.04&0.02&0.00&0.00&\multicolumn{1}{c}{n.a.}\\
\verb/sampleh4.i/&0.00&0.07&0.01&0.00&0.00&\multicolumn{1}{c}{n.a.}\\
\verb/sampleh5.i/&0.00&0.05&0.01&0.00&0.00&\multicolumn{1}{c}{n.a.}\\
\verb/sampleh6.i/&0.00&0.06&0.01&0.00&0.00&\multicolumn{1}{c}{n.a.}\\
\verb/sampleh7.i/&0.00&0.09&0.01&0.00&0.00&\multicolumn{1}{c}{n.a.}\\
\verb/sampleh8.i/&52.87&73.64&78.76&\multicolumn{1}{c}{$>1h$}&4.59&\multicolumn{1}{c}{n.a.}\\
\verb/trunc10.i/&8.81&9.06&0.08&1.66&9.15&737.37 \\
\verb/trunc7.i/&0.02&0.13&0.04&0.13&0.16&17.51 \\
\verb/tsp5.i/&0.00&0.10&0.04&0.02&0.00&\multicolumn{1}{c}{n.a.}\\
\hline
total&130.34&308.12&345.70
\end{longtable}
} 

In Table~\ref{tab:efficiency-vertex-enumeration-hard} we have collected
the data concerning the hardest of these problems.  Here we have imposed
a memory limit of 768~MB: entries marked with `mem' indicate the problems
on which this limit was exceeded.
The entries marked with `tab' in the New Polka's column indicate the problems
where New Polka ran ``out of table space'' in the conversion algorithm.

{
\renewcommand{\arraystretch}{0.91}
\small
\begin{longtable}{l......}
\caption{Efficiency on vertex enumeration: hard problems%
\label{tab:efficiency-vertex-enumeration-hard}} \\
\hline
\multicolumn{1}{c}{input} &
\multicolumn{1}{c}{PPL} &
\multicolumn{1}{c}{New Polka} &
\multicolumn{1}{c}{PolyLib} &
\multicolumn{1}{c}{cddlib} &
\multicolumn{1}{c}{lrslib} &
\multicolumn{1}{c}{pd} \\
\hline
\endhead
\hline
\endfoot
\verb/cp7.e/&\multicolumn{1}{c}{$>1h$}&\multicolumn{1}{c}{tab}&\multicolumn{1}{c}{$>1h$}&\multicolumn{1}{c}{$>1h$}&\multicolumn{1}{c}{$>1h$}&\multicolumn{1}{c}{$>1h$} \\
\verb/cyclic25_13.e/&89.94&\multicolumn{1}{c}{tab}&354.83&221.98&12.10&\multicolumn{1}{c}{n.a.}\\
\verb/cp6.i/&\multicolumn{1}{c}{$>1h$}&\multicolumn{1}{c}{tab}&\multicolumn{1}{c}{$>1h$}&\multicolumn{1}{c}{$>1h$}&\multicolumn{1}{c}{$>1h$}&\multicolumn{1}{c}{n.a.}\\
\verb/mit.i/&\multicolumn{1}{c}{$>1h$}&\multicolumn{1}{c}{tab}&\multicolumn{1}{c}{$>1h$}&950.00&2024.16&\multicolumn{1}{c}{n.a.}\\
\verb/mit288-281.i/&\multicolumn{1}{c}{mem}&\multicolumn{1}{c}{tab}&\multicolumn{1}{c}{mem}&\multicolumn{1}{c}{mem}&\multicolumn{1}{c}{$>1h$}&\multicolumn{1}{c}{ovfl} \\
\verb/mit41-16.i/&137.14&\multicolumn{1}{c}{tab}&320.37&325.65&33.89&\multicolumn{1}{c}{$>1h$} \\
\verb/mit708-9.i/&\multicolumn{1}{c}{$>1h$}&\multicolumn{1}{c}{tab}&\multicolumn{1}{c}{$>1h$}&1016.23&1927.18&\multicolumn{1}{c}{n.a.}\\
\verb/mit71-61.i/&\multicolumn{1}{c}{$>1h$}&\multicolumn{1}{c}{tab}&\multicolumn{1}{c}{mem}&\multicolumn{1}{c}{$>1h$}&\multicolumn{1}{c}{$>1h$}&\multicolumn{1}{c}{n.a.}\\
\verb/mit90-86.i/&\multicolumn{1}{c}{mem}&\multicolumn{1}{c}{tab}&\multicolumn{1}{c}{mem}&\multicolumn{1}{c}{$>1h$}&\multicolumn{1}{c}{$>1h$}&\multicolumn{1}{c}{ovfl} \\
\end{longtable}
} 

Another possibility of evaluating the performance of the Parma Polyhedra
Library on a standard problem with standard data is offered by linear
programming, which is the paradigm upon which several approaches
to analysis and verification
(such as, e.g., \cite{SankaranarayananCSM06,SankaranarayananSM05})
rest upon.  This requires either a version of the simplex based on
exact arithmetic, or, in case a classical floating-point
implementation is used, it forces to validate the obtained result
with some alternative methods.
The \emph{MathSAT}\footnote{\url{http://mathsat.itc.it/}.} decision procedure
\cite{BozzanoBCJRSS05},
which is applicable to the formal verification of infinite state systems
(such as timed and hybrid systems), is based on a version of
the \emph{Cassowary Constraint Solving Toolkit} \cite{BadrosBS01},
modified so as to use
exact arithmetic instead of floating-point numbers.
Moreover, the algorithm employed by MathSAT requires incremental
satisfiability checks: a set of constraints is added and satisfiability
is checked, more constraints are added and satisfiability is re-checked,
and so forth.
We have thus measured the efficiency of the PPL's incremental constraint
solver by comparison with the version of Cassowary used in MathSAT and with
the \emph{Wallaroo Linear Constraint Solving
Library},\footnote{\url{http://sourceforge.net/projects/wallaroo/}.}
another descendant of Cassowary.
The benchmarks we used are quite standard in the linear programming community:
they come from the `\verb+lp+' directory of
\emph{NetLib}.\footnote{\url{http://www.netlib.org/lp/index.html}.}
The solution times, in seconds, obtained for the problem of adding one
constraint at a time, checking for satisfiability at each step, are given in
Table~\ref{tab:efficiency-simplex-incremental-satisfiability}.

{
\renewcommand{\arraystretch}{0.91}
\small
\begin{longtable}{l......}
\caption{Efficiency of the simplex solver on incremental satisfiability checking%
\label{tab:efficiency-simplex-incremental-satisfiability}} \\
\hline
\multicolumn{1}{c}{input} &
\multicolumn{1}{c}{PPL} &
\multicolumn{1}{c}{Wallaroo} &
\multicolumn{1}{c}{Cassowary/MathSAT} \\
\hline
\endhead
\hline
\endfoot
\verb/adlittle.mps/ &  0.33 &  1.46 &  1.51 \\
\verb/afiro.mps/    &  0.02 &  0.05 &  0.07 \\
\verb/blend.mps/    & 13.45 &  5.40 &  8.23 \\
\verb/boeing1.mps/  & 47.28 & 87.80 & 75.48 \\
\verb/boeing2.mps/  &  2.32 & 10.58 & 14.67 \\
\verb/kb2.mps/      &  0.11 &  0.30 &  0.46 \\
\verb/sc105.mps/    &  0.48 & 10.95 &  7.23 \\
\verb/sc50a.mps/    &  0.05 &  0.64 &  0.56 \\
\verb/sc50b.mps/    &  0.06 &  0.70 &  0.94\\
\hline
total                & 64.10 &117.88 &109.15
\end{longtable}
} 

\section{Development Plans}
\label{sec:development-plans}

In this section we briefly review the short- and mid-term development
plans we have for the library.  We deliberately omit all long-term
projects: for all those we mention here, code ---whether in the form of a
prototype or as a proof-of-concept exercise--- has already been developed
that proves the feasibility of the proposal.

\subsection{More Abstractions}

\paragraph*{Intervals and Bounding Boxes}
An important numerical domain is the domain of \emph{bounding boxes}:
these are representable by means of finite set of \emph{intervals} or
be seen as finite conjunctions of constraints of the form $\pm v_i \leq d$
or $\pm v_i < d$.  Despite the fact that bounding boxes have been one of
the first abstract domains ever proposed \cite{CousotC76} and that they
have been implemented and reimplemented dozens of times, no freely available
implementation is really suitable for the purposes of abstract interpretation.
In fact, the available interval libraries either lack support
for non-closed intervals (so that they are unable to represent
constraints of the form $\pm v_i < d$), or they do not provide the right
support for approximation in the sense of partial correctness
(e.g., division by an interval containing zero gives rise to a run-time error
instead of giving an interval containing the result under the assumption
that the concrete division being approximated was not a division by zero), or
they disregard rounding errors and are therefore unsafe.
We are thus working at a complete implementation of bounding boxes based
on intervals. Such intervals are parametric on a number of features:
they support open as well as closed boundaries;
boundaries can be chosen within one of the number families mentioned
in Section~\ref{sec:bounded-difference-shapes} (when boundaries are floating
point numbers, rounding is of course controlled to maintain soundness);
independently from the type of the boundaries, both plain intervals of real
numbers and intervals subject to generic \emph{restrictions} are supported.
This notion of restriction can be instantiated to obtain intervals of integer
numbers, \emph{modulo intervals} \cite{NakanishiF01,NakanishiJPF99},
and generalizations of the latter providing more precise
information.\footnote{An implementation of these interval families
is already available in the PPL's public CVS repository.}

\paragraph*{Grid-Polyhedra}
An interesting line of development consists in the combination of the grids
domain with the several polyhedral domains provided by the PPL:
not only the $\mathbb{Z}$-polyhedra domain \cite{Ancourt91th},
but also many variations such as grid-polyhedra, grid-octagon,
grid-bounded-difference, grid-interval domains
(not to mention their powersets).

\paragraph*{Polynomial Equalities and Inequalities}
The work in \cite{BagnaraR-CZ05} proved the feasibility of representing
systems of polynomial inequalities of bounded degree by encoding them
into convex polyhedra.  The prototype implementation used for the
experimental evaluation presented in \cite{BagnaraR-CZ05} is being
turned into a complete abstract domain and will be incorporated into
the PPL.

\subsection{More Language Interfaces}

The current version of the PPL only offers C and Prolog interfaces for
(C and NNC) polyhedra and LP problems.  It would not be difficult to
add, along the same lines, interfaces for all the other abstractions.
It can be done, rather quickly, mostly as a ``copy and paste'' exercise.
Instead of following that route (which would imply substantial code
duplication and an unaffordable maintenance burden), we are working
at an automatic way of obtaining these interfaces out of a few ``templates.''
As part of this ongoing effort, we are extending the set of available
\emph{direct} interfaces\footnote{As opposed to the \emph{indirect}
interfaces that can be obtained by passing through the C interface.}
with Java and Objective Caml.\footnote{Initial versions of these interfaces
are already available in the PPL's public CVS repository.}
Finally there are plans to develop an interface that allows
to use the PPL's numerical abstractions within
\emph{Mathematica}.\footnote{\url{http://www.wolfram.com/}.}

\subsection{Other Features}

Other features on the horizon of the Parma Polyhedra Library include
the inclusion of:
bidirectional serialization functions for all the supported abstractions;
the ask-and-tell generic construction of \cite{Bagnara97th};%
\footnote{Preliminary support for both these features is already available
in the PPL's public CVS repository.}
and the extrapolation operators defined in \cite{HenzingerH95}
and \cite{HenzingerPW01};
We also plan to add support for more complexity-throttling techniques
such as:
breaking down the set of the variables of interests into ``packs''
of manageable size \cite{BlanchetCCFMMMR03,CousotCFMMMR05,VenetB04};
support for Cartesian factoring as defined in \cite{HalbwachsMP-V03};
and the limitation of the number of constraints and generators
and/or the size of coefficients in the representation
of polyhedra \cite{Frehse05}.

\section{Discussion}
\label{sec:discussion}

In this paper, we have presented the Parma Polyhedra Library, a library
of numerical abstractions especially targeted to applications in the
field of analysis and verification of software and hardware systems.
We have illustrated the general philosophy that is behind the design
of the library, its main features, examples highlighting the advantages
offered to client applications, and the avenue we have prepared for
future developments.

The Parma Polyhedra Library is being used on several applications
in the field of verification of hardware and software systems.
It has been used for the verification of properties of oscillator circuits
\cite{FrehseKR06,FrehseKRM05};
to verify the soundness of \emph{batch workflow networks}
(a kind of Petri nets used in workflow management) \cite{vanHeeOSV06};
in the field of safety analysis of continuous and hybrid systems
to overapproximate the systems of linear differential equations
expressing the dynamics of hybrid automata
\cite{DoyenHR05,Frehse04,FrehseHK04,SankaranarayananSM06} and,
in particular, the PPL is used in PHAVer, an innovative tool for the
verification of such systems \cite{Frehse05}.
The PPL is also used: in a version of \emph{TVLA} (3-Valued Logic Analysis
Engine, \url{http://www.cs.tau.ac.il/~tvla/}), a system for the verification
of properties of arrays and heap-allocated data \cite{GopanRS05};
in \emph{iCSSV} (interprocedural C String Static
Verifier), a tool for verifying the safety of string operations in C programs
\cite{Ellenbogen04th};
and in a static analyzer
for \emph{gated data dependence graphs}, an intermediate representation
for optimizing compilation \cite{HymansU04}.  This analyzer employs,
in particular, the precise widening operator and the widening with tokens
technique introduced in \cite{BagnaraHRZ03,BagnaraHRZ05SCP}.
In \cite{SankaranarayananCSM06} the PPL is used to derive invariant linear
equalities and inequalities for a subset of the C language;
it is used in \emph{StInG} \cite{SankaranarayananSM04} and \emph{LPInv}
\cite{SankaranarayananSM05}, two systems for the analysis of transition
systems;
it is used for the model-checking of reconfigurable hybrid
systems \cite{SongCR05};
it is used in a static analysis tool for x86 binaries
that automatically identifies instructions
that can be used to redirect control flow, thus constituting
vulnerabilities that can be exploited in order to bypass
intrusion detection systems \cite{KruegelKMRV05};
it is also used to represent and validate real-time systems' constraints
and behaviors \cite{DooseM05}
and to automatically derive the \emph{argument size relations}
that are needed for termination analysis of Prolog programs
\cite{MesnardB05TPLP}.

In conclusion, even though the library is still not mature and
functionally complete, it already offers a combination of
functionality, reliability, usability and performance that is not
matched by similar, freely available libraries.  Moreover, since the
PPL is free software and distributed under the terms of the GNU
General Public License (GPL), and due to the presence of extensive
documentation, the library can already be regarded as an important
contribution secured to the community.

For the most up-to-date information, documentation and downloads
and to follow the development work, the reader is referred to the
Parma Polyhedra Library site at \url{http://www.cs.unipr.it/ppl/}.

\bigskip
\noindent
{\bfseries Acknowledgments.}
We would like to express our gratitude and appreciation to all the
present and past developers of the Parma Polyhedra Library:
\italian{Irene Bacchi},
\italian{Abramo Bagnara},
\italian{Danilo Bonardi},
\italian{Sara Bonini},
\italian{Andrea Cimino},
Katy Dobson,
\italian{Giordano Fracasso},
\italian{Maximiliano Marchesi},
\italian{Elena Mazzi},
David Merchat,
Matthew Mundell,
\italian{Andrea Pescetti},
\italian{Barbara Quartieri},
\italian{Elisa Ricci},
Enric Rodr\'\i guez-Carbonell,
\italian{Angela Stazzone},
\italian{Fabio Trabucchi},
\italian{Claudio Trento},
\italian{Alessandro Zaccagnini},
\italian{Tatiana Zolo}.
Thanks also
to Aaron Bradley, for contributing to the project his
\emph{Mathematica} interface,
to Goran Frehse, for contributing his code to limit the complexity
of polyhedra,
and to all the users of the library that provided us with
helpful feedback.


\newcommand{\noopsort}[1]{}\hyphenation{ Ba-gna-ra Bie-li-ko-va Bruy-noo-ghe
  Common-Loops DeMich-iel Dober-kat Di-par-ti-men-to Er-vier Fa-la-schi
  Fell-eisen Gam-ma Gem-Stone Glan-ville Gold-in Goos-sens Graph-Trace
  Grim-shaw Her-men-e-gil-do Hoeks-ma Hor-o-witz Kam-i-ko Kenn-e-dy Kess-ler
  Lisp-edit Lu-ba-chev-sky Ma-te-ma-ti-ca Nich-o-las Obern-dorf Ohsen-doth
  Par-log Para-sight Pega-Sys Pren-tice Pu-ru-sho-tha-man Ra-guid-eau Rich-ard
  Roe-ver Ros-en-krantz Ru-dolph SIG-OA SIG-PLAN SIG-SOFT SMALL-TALK Schee-vel
  Schlotz-hauer Schwartz-bach Sieg-fried Small-talk Spring-er Stroh-meier
  Thing-Lab Zhong-xiu Zac-ca-gni-ni Zaf-fa-nel-la Zo-lo }

\end{document}